\documentclass[twocolumn,prl,10pt,aps,floatfix]{revtex4-1}
\usepackage{graphicx}
\usepackage{amsmath}
\usepackage[colorlinks,linkcolor=blue,citecolor=blue,urlcolor=blue]{hyperref}
\usepackage{color}

\begin{document}
\title{Ballistic magnetic thermal transport coupled to phonons}
\author{P. Zavitsanos$^{1}$ and X. Zotos$^{1,2,3}$}
\affiliation{$^1$Department of Physics,
University of Crete, 70013 Heraklion, Greece}
\affiliation{$^2$Foundation for Research and Technology - Hellas, 71110
Heraklion, Greece}
\affiliation{
$^3$Leibniz Institute for Solid State and Materials Research IFW Dresden,
01171 Dresden, Germany}

\date{\today}

\begin{abstract}
Motivated by thermal conductivity experiments
in spin chain compounds, we propose a phenomenological model to account for a 
ballistic magnetic transport coupled to a diffusive phononic one,
along the line of the seminal two-temperature diffusive 
transport Sanders-Walton model.
Although the expression for the effective thermal conductivity
is identical to that of Sanders-Walton, 
the interpretation is entirely different, as
the "magnetic conductivity" is replaced by an "effective transfer
conductivity" between the magnetic and phononic component.
This model also reveals the fascinating possibility of visualizing
the ballistic character of magnetic transport,
for appropriately chosen material parameters,
as a two peak counter-propagating feature in the phononic temperature.
It is also appropriate for the analysis of any
thermal transport experiment involving a diffusive component coupled to a 
ballistic one.

\end{abstract}
\maketitle

The thermal transport by magnetic excitations has been extensively studied 
over the last few years \cite{hess} and it was established as a novel 
thermal conduction mechanism besides 
the well known electronic and phononic ones. 
In particular, it was studied in one dimensional quantum 
magnets \cite{ballistic} where it was shown that, 
in compounds accurately described 
by the Heisenberg spin-1/2 chain model, there is a ballistic magnetic 
component of thermal transport \cite{znp} interacting with the phononic one. 

Besides steady state studies of the thermal conductivity, the flash 
method\cite{parker} 
provides information on the dynamic (in time) propagation of heat 
and in particular on the interaction between 
the magnetic and phononic components \cite{flash}.
These studies could  provide a playground for confronting experiments  
to recent theoretical developments on 
the far-out of equilibrium dynamics in integrable 
spin Hamiltonians \cite{ghd}. 

However, the magnetic excitations in a quantum magnet 
are interacting with the phonons  
and disentangling their contribution in the total thermal 
conductivity is an important issue.
A minimal phenomenological framework of thermal conduction in a diffusive 
magnon plus phonon system was proposed by Sanders and Walton (SW) 
\cite{sw} and it has 
become the standard model for analyzing magnetic thermal 
transport experiments \cite{ott}.
In the SW model the magnetic subsystem is assumed diffusive so it is 
interesting to revisit this model in the case of a ballistic magnetic component
more appropriate for the spin-1/2 Heisenberg chain compounds and not only. 
In the following we will try to 
highlight the differences in the expected effective thermal conductivity 
and thermal pulse propagation between the two models, 
namely the two-temperature SW diffusion model and the present 
advection-diffusion one.

In SW the starting relation is the equilibration in time of the 
phonon ($T_{p}$) and magnetic ($T_m$) temperature difference,
\begin{equation}
\frac{\partial \Delta T}{\partial t}=-\frac{\Delta T}{\tau},
~~~\Delta T=T_p-T_m
\end{equation}
\noindent
where $\tau$ is a characteristic phenomenological relaxation time. 
This basic relation is also satisfied by a system of 
individual contributions,

\begin{eqnarray}
\frac{\partial T_p}{\partial t}&=&\frac{c_m}{C} \frac{T_m-T_p}{\tau}
\nonumber\\
\frac{\partial T_+}{\partial t}&=&\frac{c_p}{C} \frac{T_p-T_+}{\tau}
\nonumber\\
\frac{\partial T_-}{\partial t}&=&\frac{c_p}{C} \frac{T_p-T_-}{\tau},
\end{eqnarray}

\noindent
from right (left) moving magnetic carriers  
with temperature $T_{\pm}$ ($T_m=(T_+ + T_-)/2$) and
phonons at temperarure $T_p$. Here, $c_{\pm}$ are the corresponding magnetic 
($c_m=c_+ + c_-$) and $c_p$ phonon specific heats, $C=c_p + c_m$ the 
total specific heat.

\noindent
These relations are for space independent temperature profiles.
We can extend them to a space dependent energy diffusion equation  
for the phonon subsystem and two advection equations for the ballistic 
magnetic system, 

\begin{eqnarray}
&&\frac{\partial \epsilon_p}{\partial t}=D \frac{\partial^2 \epsilon_p}
{\partial x^2} + \frac{c_p c_m}{C}\frac{T_m - T_p}{\tau} 
\nonumber\\
&&\frac{\partial \epsilon_{\pm}}{\partial t} 
\pm v\frac{\partial \epsilon_{\pm}}{\partial x} =
\frac{c_p c_{\pm}}{C}\frac{T_p - T_{\pm}}{\tau}. 
\end{eqnarray}

\noindent
$D$ is the phonon diffusion constant, $v$ the characteristic velocity 
of magnetic excitations, $\epsilon_p$ the phonon energy density, 
$\epsilon_{\pm}$ the magnetic ones and $\delta \epsilon_{p,\pm}=c_{p,\pm} 
\delta T_{p,\pm}$. To have a concrete model in mind, for the low energy 
spinon gas in the 1D spin-1/2 Heisenberg chain with energy dispersion 
$\epsilon_{\rm spinon}\simeq v|p|$, 
$\epsilon_{\pm}=\frac{\pi}{12}\frac{T_{\pm}^2}{v}$, 
$c_{\pm}=\frac{\pi}{6}\frac{T_{\pm}}{v}$ and $v$ the spinon velocity. 
Here and in the following, we consider small deviations from thermal 
equilibrium and take the specific heats independent of temperature.

Furthermore, the total energy current density $Q=Q_p + Q_m$ is given by 
 the phonon $Q_p$ and magnetic $Q_m$ energy currents, 
\begin{eqnarray}
Q_p&=&-\kappa_p \frac{\partial T_p}{\partial x}
\nonumber\\
Q_m&=&v\epsilon_+ - v\epsilon_-,
\end{eqnarray}

\noindent
with $\kappa_p=c_p D$ the phonon thermal conductivity.
Reverting back to temperature dependent equations,

\begin{eqnarray}
&&\frac{\partial T_p}{\partial t}=D \frac{\partial^2 T_p}
{\partial x^2} + \frac{c_m}{C}\frac{T_m - T_p}{\tau} 
\nonumber\\
&&\frac{\partial T_{\pm}}{\partial t} 
\pm v\frac{\partial T_{\pm}}{\partial x} =
\frac{c_p}{C}\frac{T_p - T_{\pm}}{\tau} 
\nonumber\\
&&Q=-\kappa_p \frac{\partial T_p}{\partial x}
+v\frac{c_m}{2}(T_+ - T_-).
\label{advdiff}
\end{eqnarray}

We will first consider the effective thermal conductivity of a system 
$-L/2 < x < L/2$ in a steady state with only a phonon energy 
currrent $Q|_{x=\pm L/2} =Q_p$ at its borders 
and no magnetic current $(T_+ - T_-)|_{x=\pm L/2}=0$.
The steady state equations become,

\begin{eqnarray}
&&v\frac{\partial T_m}{\partial x}=-\frac{c_p}{C} \frac{\Delta T_m}{\tau}
,~~~ \Delta T_m =\frac{T_+ - T_-}{2}
\nonumber\\
&&D \frac{\partial^2 T_p}{\partial x^2} +\frac{c_m}{C}\frac{T_m-T_p}{\tau}=0,
\label{sst}
\end{eqnarray}
\begin{equation}
Q=-\kappa_p\frac{\partial T_p}{\partial x} + v c_m \Delta T_m.
\label{q}
\end{equation}

\noindent
Solving (\ref{q}) for $\Delta T_m$ we obtain,

\begin{eqnarray}
\Delta T_m&=&\frac{1}{v c_m}(Q+\kappa_p \frac{\partial T_p}{\partial x})
\nonumber\\
\frac{\partial T_m}{\partial x}&=&-\frac{1}{\tilde{\kappa}_m} 
(Q+\kappa_p \frac{\partial T_p}{\partial x}), 
\end{eqnarray}

\noindent
where ${\tilde\kappa}_m=(c_m/c_p)Cv^2 \tau$ 
is an effective {\it transfer thermal conductivity}.
We solve this equation by, 
(i) assuming that $c_p, c_m$ are 
temperature independent (which strictly speaking is not the case) and
(ii) taking the boundary condition $T_p(x=0)=T_m(x=0)=T_0$. 
We find,
\begin{equation}
T_m=T_0 - \frac{\kappa_p}{\tilde{\kappa}_m}(T_p-T_0)-
\frac{1}{\tilde{\kappa}_m}Q x 
\label{tm}
\end{equation}

\noindent
and by substituting in (\ref{sst}),

\begin{eqnarray}
&&\frac{\partial ^2(T_p - T_0)}{\partial x^2} 
- A^2 (T_p - T_0) - \frac{A^2}{\kappa_t}Qx=0
\nonumber\\
&&A^2 =\frac{c_pc_m}{C\tau}\cdot\frac{\kappa_t}{\kappa_p{\tilde\kappa}_m},~~
~\kappa_t=\kappa_p+{\tilde{\kappa}}_m.
\label{ssp}
\end{eqnarray}
\noindent
The solution of (\ref{ssp}) with the boundary condition 
$\partial (T_p-T_0)/\partial x=-Q/\kappa_p$ gives the phonon temparature 
profile,
\begin{equation}
T_p=T_0-\frac{x}{\kappa_t}Q-
\frac{\tilde{\kappa}_m}{\kappa_t\kappa_p}\frac{\sinh Ax}{A\cosh AL/2}Q,
\label{tp}
\end{equation}

\noindent
and (\ref{tm}) the magnetic temperature one.

The effective thermal conductivity obtained from 
$\kappa_{eff}=-QL/\Delta T_p$ \cite{sw} is given by,
\begin{eqnarray}
\kappa_{eff}&=&\kappa_t\Big(1+\frac{\tilde{\kappa}_m}{\kappa_p}
\frac{\tanh(AL/2)}{AL/2}\Big)^{-1},
\nonumber\\
\kappa_{eff}&\sim& \kappa_p,~~~
AL \rightarrow 0
\nonumber\\
\kappa_{eff}&\sim& \kappa_p+\tilde{\kappa}_m=\kappa_t,~~~
AL \rightarrow \infty.
\label{kappaeff}
\end{eqnarray}

The above relations are identical to those of the SW two-temperature 
model with the replacement of the magnetic conductivity by the 
effective magnetic transfer one $\tilde{\kappa}_m=(c_m/c_p)Cv^2\tau$.

\bigskip
Next, we will discuss the time dependent evolution of phonon and magnetic
temperature profiles (\ref{advdiff}) 
that can be probed by the flash method \cite{parker,flash}.
Considering an open system, $0< x<L$, 
we set as zero energy current boundary conditions,
$\partial T_p/\partial x|_{x=0,L}=0$ and $T_+ = T_-|_{x=0,L}$.
We seek solutions of the form,
\begin{eqnarray}
T_p&=&\frac{a_0}{2}+\sum_{n=1}^{+\infty} a_n \cos q_n x,~~~q_n=\frac{\pi n}{L}
\nonumber\\
T_{\pm}&=&\frac{b_0}{2}+\sum_{n=1}^{+\infty} b_n \cos q_n x
\pm c_n \sin q_n x.
\label{tppm}
\end{eqnarray}

\noindent
By substituting (\ref{tppm}) in (\ref{advdiff}),
we obtain the time dependence of $a_n, b_n, c_n$,
\begin{eqnarray}
&&{\dot a}_0+\frac{c_m}{C\tau}(a_0-b_0)=0
\nonumber\\
&&{\dot b}_0+\frac{c_p}{C\tau}(b_0-a_0)=0
\end{eqnarray}
\noindent
with solution, 
\begin{eqnarray}
a_0(t)=a_0(0)-\frac{c_m}{C}(a_0-b_0)|_{t=0}(1-e^{-t/\tau})
\nonumber\\
b_0(t)=b_0(0)-\frac{c_p}{C}(b_0-a_0)|_{t=0}(1-e^{-t/\tau}).
\end{eqnarray}
\noindent
For finite wavevector $q_n$, 
\begin{eqnarray}
&&{\dot a}_n+(D q_n^2) a_n + \bar{c}_m(a_n-b_n)=0
\nonumber\\
&&{\dot b}_n+(v q_n) c_n + \bar{c}_p(b_n-a_n)=0
\nonumber\\
&&{\dot c}_n-(v q_n) b_n + \bar{c}_pc_n=0,
\label{dotn}
\end{eqnarray}
\noindent
where $\bar{c}_p=c_p/(C\tau),~~~\bar{c}_m=c_m/(C\tau)$ are $O(1/\tau)$.
Solutions of the form $e^{\lambda t}$ are obtained by  
solving the characteristic 3rd order polynomial equation,
\begin{eqnarray} 
&&{\tilde \lambda}^3+
(Dq_n^2+\bar{c}_m-\bar{c}_p){\tilde \lambda}^2+ 
((vq_n)^2-\bar{c}_p\bar{c}_m){\tilde \lambda}+
\nonumber\\
&&(Dq_n^2+\bar{c}_m-\bar{c}_p)(vq_n)^2=0
\nonumber\\
&&\lambda={\tilde \lambda}-\bar{c}_p.
\label{roots}
\end{eqnarray} 
\noindent
The roots of this polynomial, although 
known \cite{cardano}, are not physically transparent.
To proceed, we will make the physical assumption that 
the relaxation time $\tau$ is the shortest scale 
in the problem. 
Thus we expect two roots of the order $1/\tau$ presenting the relaxation 
of the magnetic excitations and one root 
of the order of the diffusion constant. 
Once the roots determined, the constants $\alpha_i,\beta_i,\gamma_i$ 
of the time evolutions,
$a_n(t)=\sum_{i=1}^3 \alpha_ie^{\lambda_it},~~
b_n(t)=\sum_{i=1}^3 \beta_ie^{\lambda_it},~~
c_n(t)=\sum_{i=1}^3 \gamma_ie^{\lambda_it}$ 
are evaluated from (\ref{roots}) and the initial conditions
$a_n(t=0),b_n(t=0),c_n(t=0)$. 

As an example, in the recent experiment \cite{flash} 
the relevant quantities for SrCuO$_2$ are, $\kappa\simeq 50$~W/Km,
$\kappa_p\simeq 8$~W/Km, 
$c_p \simeq 2.8 \cdot 10^6$~J/Km$^3$,~~$c_m \simeq 3\cdot 10^4$~J/Km$^3$
($c_p \sim 100 c_m$), 
$D =\kappa_p/c_p\sim 3\cdot 10^{-6}$ m$^2$/s,  
$v \simeq 2\cdot 10^4$~m/s , $\tau\sim O(10^{-12}$ s), which 
gives ${\tilde \kappa}_m\sim O(10$~W/Km).
The largest uncertainty in these parameters is in the relaxation 
time $\tau$.
For a typical sample of length $L\sim 1$~mm these imply
three well separated characteristic time scales, $D/L^2 \sim 3$ s$^{-1}$, 
$(v/L)\sim 2\cdot 10^7$ s$^{-1}$ and $1/\tau \sim 10^{12}$ s$^{-1}$. 
For these parameters, $AL\sim O(10^5)$ so that $\kappa_{eff}\simeq
\kappa_p+\tilde{\kappa}_m$.

Furthermore, for these experimental values, keeping the 
dominant terms in (\ref{roots}) (e.g. dropping the 1st term, 
taking $Dq_n^2\tau\rightarrow 0$ and 
$c_m << c_p$) we obtain ,

\begin{eqnarray}
\lambda_2&\simeq& -\frac{1}{\tau}
+\frac{(vq_n)^2}{{\bar c}_m}
\nonumber\\
\lambda_3&\simeq& -\frac{c_p}{C\tau} 
-\frac{(vq_n)^2}{{\bar c}_m}.
\label{root23exp}
\end{eqnarray}

\noindent
Next, taking $\lambda \sim O(\epsilon)$, $\epsilon\tau << 1$, 
substituting in (\ref{roots}) $\tilde \lambda =\epsilon +\bar{c}_p$ 
and keeping 1st order terms in $\epsilon$, we find,
\begin{equation} 
\lambda_1 \simeq -(D+\frac{\bar{c}_m}{\bar{c}_p}\cdot \frac{v^2}{\bar{c}_p})
q_n^2.
\label{root1}
\end{equation} 
\noindent
This relation 
implies a total diffusion constant composed of a phononic component 
$D\simeq 3\cdot 10^{-6}$~m$^2$/s enhanced 
by the ballistic magnetic component 
$\frac{\bar{c}_m}{\bar{c}_p}\cdot \frac{v^2}{\bar{c}_p}
=\frac{c_m}{c_p}\cdot \frac{v^2 C \tau}{c_p}
\sim O(10^{-6}$~m$^2$/s). Consistently, 
multiplying (\ref{root1}) by $c_p$ 
we recover (\ref{kappaeff}) with the second term corresponding 
to the effective magnetic transfer ${\tilde\kappa}_m$.
In this limit, $\alpha_1\simeq\beta_1\simeq-\beta_2\simeq a_n(t=0)$ and 
$\alpha_3\simeq\beta_3\simeq\gamma_i\simeq 0$.
Thus, a flash method experiment that probes the long time behavior of the 
temperature profile, when analyzed in terms of 
a diffusion equation, gives an effective diffusion constant with a 
phononic and magnetic contibution. 

Whether we have three real roots or one real and two complex
conjugate ones, indicating oscillatory behavior, depends on the sign 
of the discriminant in the roots $\lambda_{2,3}$,
\begin{equation}
\Delta=(\bar{c}_m)^2-4(vq_n)^2.
\label{discr}
\end{equation}
\noindent
Assuming the experimental values quoted above 
and tuning the relaxation time, we find oscillatory behavior (complex roots) 
of the magnetic component relaxation for a window 
of $\tau$ less than about $\sim 10^{-9}$ sec giving, 
\begin{equation} 
\lambda_{2,3} \sim -\frac{1}{\tau} \pm i(vq_n).
\end{equation} 
Thus, the typical scenario emerging is that, 
within a time $\tau$ there is equilibration of the magnetic temperature profile 
to the phononic one, eventually with an oscillatory behavior, followed by 
diffusive propagation of the combined system with an effective diffusion 
constant. 

Note that,  
in a flash experiment the heat is deposited on the phonon system and then 
it relaxes to the magnetic system, propagating coupled thereafter.
And, at any instant, it is the phonon temperature that it is probed. 
It would be particularly interesting if for some material parameters we can 
visualize the ballistic propagation of the magnetic temperature profile.
It is clear that at any time the magnetic profile separates in two 
wavefronts propagating left-right while relaxing to the phonon bath. 
The question is whether this two-bump profile can be manifested on the phonon 
temperature profile. This would be a telltale sign of ballistic 
propagation. To realize such a scenario within the diffusion time, 
it is favorable to have a magnetic system with large specific heat that will 
rapidly propagate, thus large velocity and relaxation time comparable 
to the diffusion time. 

\begin{figure}[!h]
\begin{center}
\includegraphics[width=1\linewidth, angle=0]{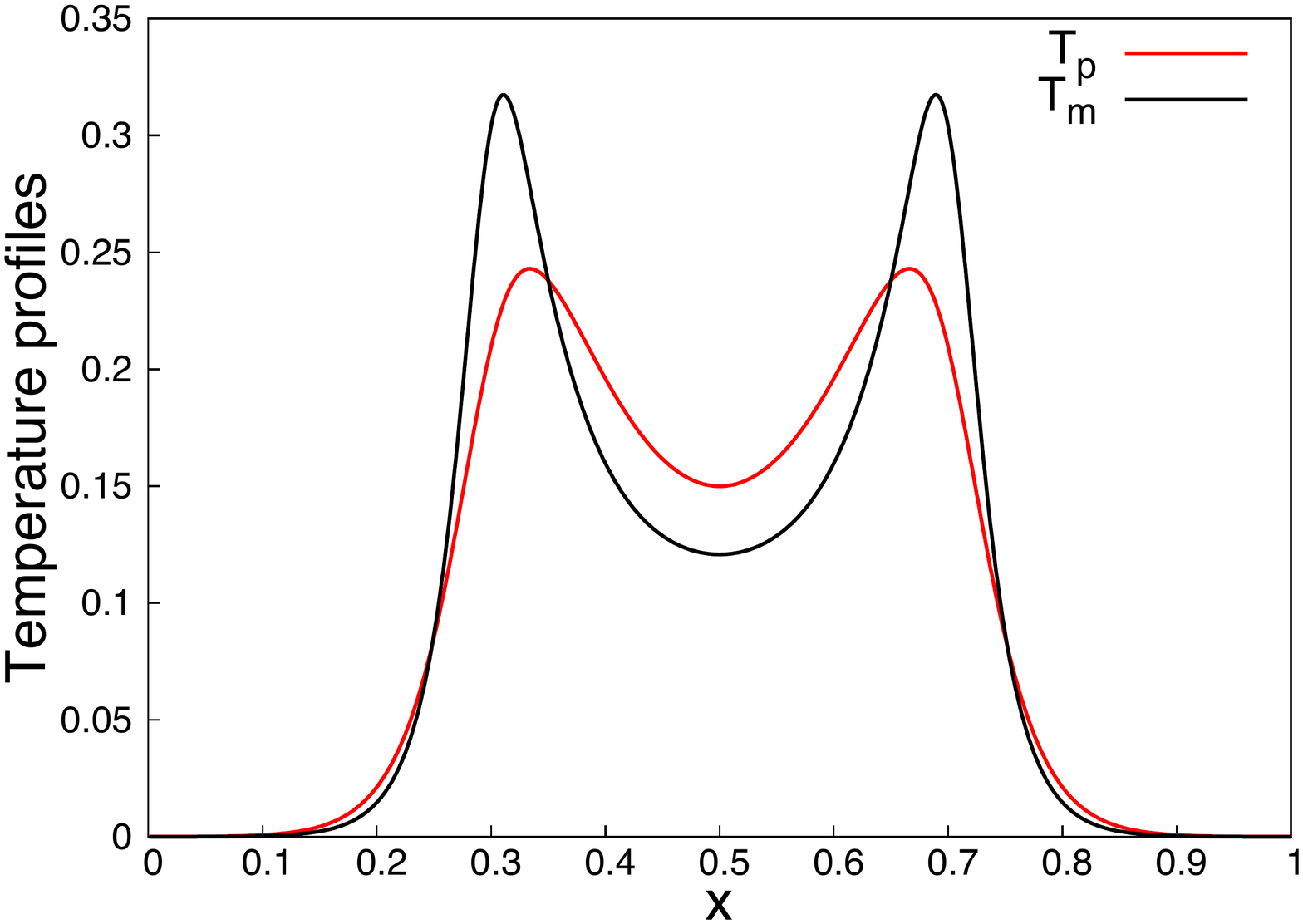}
\caption{Temperature profiles at $t=0.01s$ for 
$L=1mm,~~D_p/L^2=1s^{-1},~~c_p/C=0.1,~~
c_m/C=0.9,~~v=20 m/s,~~\tau=10^{-3}s$.
Initial temperature profiles, $T_p=(40/\sqrt{\pi})e^{-(40(x-1/2))^2},~~
T_m(t=0)=0$.}
\label{fig1}
\end{center}
\end{figure}

It is clear that the parameter space of magnetic materials to 
search for such a behavior is very extended. 
Here we will discuss just an arbitrary example with model parameters 
indicated in Fig.\ref{fig1}, where we observe such an evolution 
by numerical integration of (\ref{advdiff}).
We first take as initial 
phonon temperature profile a narrow Gaussian at the center of the sample.
Note, that for these parameters, the discriminant ({\ref{discr}) 
is large and negative, indicative of a strong oscillatory behavior. 

\begin{figure}[!h]
\begin{center}
\includegraphics[width=1\linewidth, angle=0]{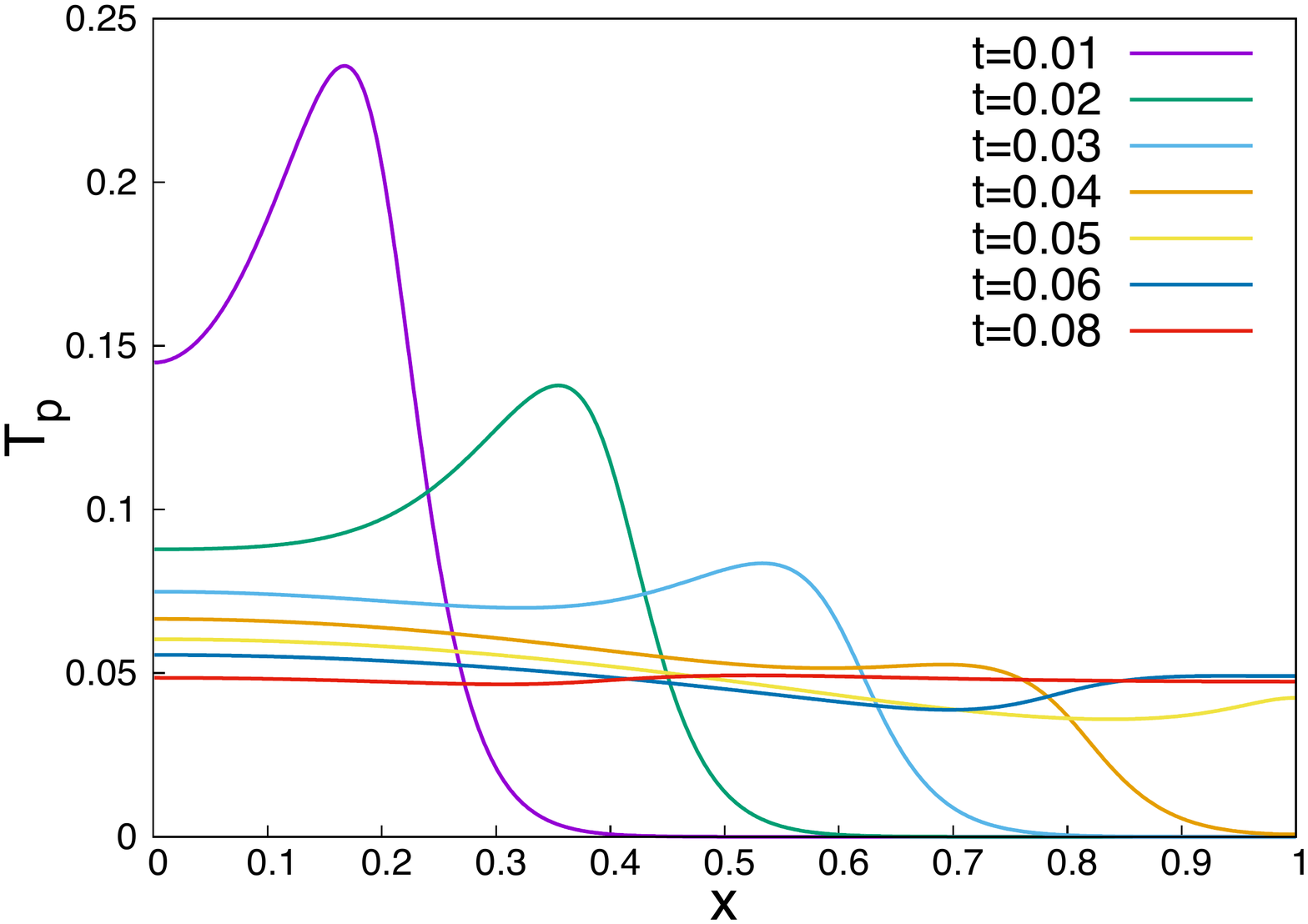}
\caption{Phonon temperature profiles for the same parameters as in 
Fig.\ref{fig1}.
Initial temperature profiles 
$T_p(t=0)=(40/\sqrt{\pi})e^{-(40x)^2},~~T_m(t=0)=0$.}
\label{fig2}
\end{center}
\end{figure}

\begin{figure}[!h]
\begin{center}
\includegraphics[width=1\linewidth, angle=0]{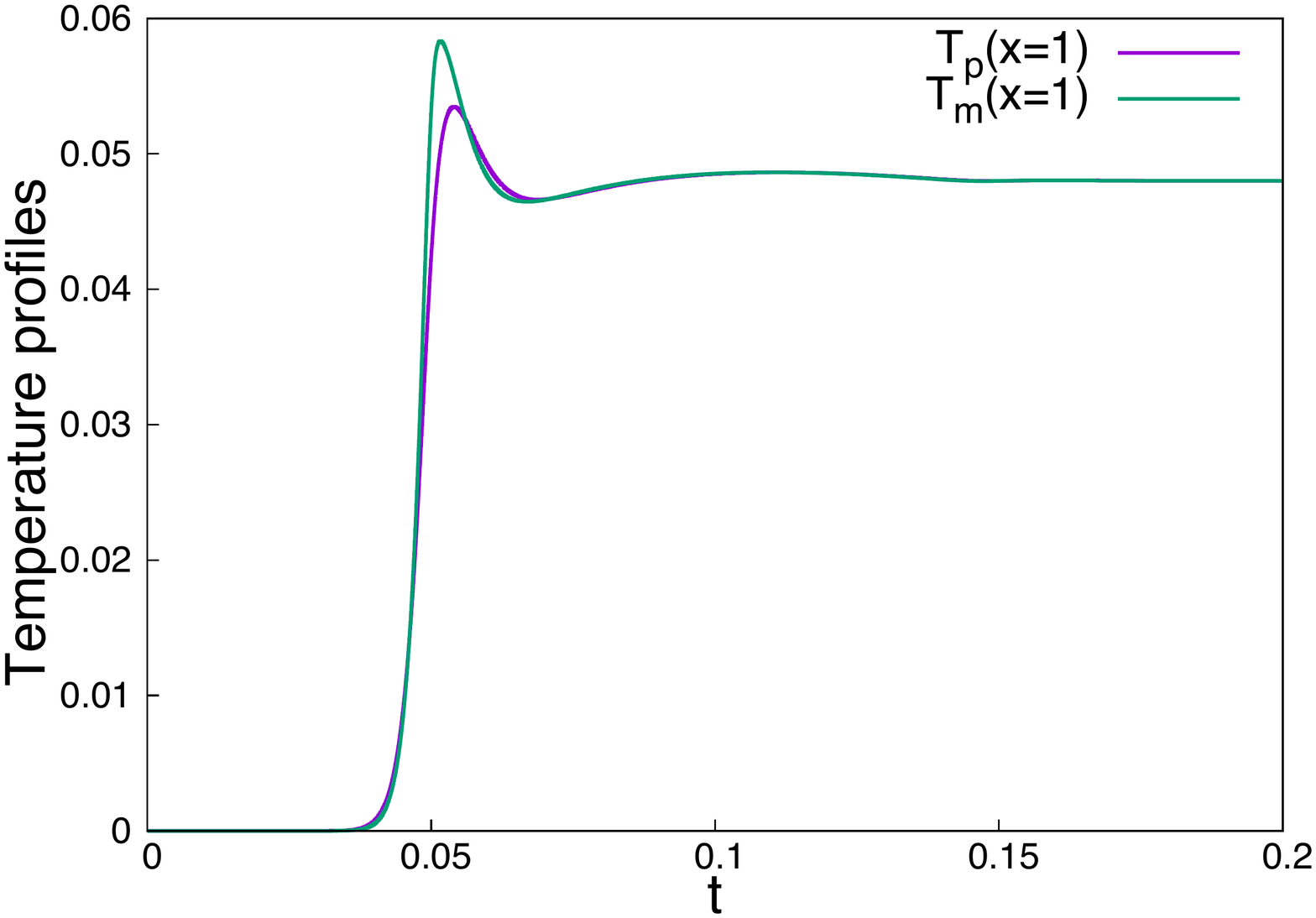}
\caption{Temperature profiles at $x/L=1$ for 
the same parameters as in Fig.\ref{fig1}.
Initial temperature profiles 
$T_p=(40/\sqrt{\pi})e^{-(40x))^2},~~T_m(t=0)=0$.}
\label{fig3}
\end{center}
\end{figure}

Next, for an initial heat pulse at the left-side of the sample deposited 
at the phonon subsystem as in a flash experiment,
we observe in Fig.\ref{fig2} the propagation of a heat wave, very unlike to 
a diffusive one, due to the hybridization of the phonon and magnetic 
components. It results to a characteristic non-monotonic 
behavior at the right side of the sample as shown in Fig.\ref{fig3}.
Certainly, further study is required to establish the range of 
parameters where such a singular temperature evolution can be observed.

In conclusion, to appropriatelly describe the ballistic thermal transport 
of magnetic excitations coupled to phonons we have introduced a 
phenomenological advection-diffusion model. This model shows qualitatively 
different behavior to the SW two-temperature diffusion model. 
In particular, it implies a 
greatly enhanced effective diffusion constant 
for the material parameters of a recent dynamic heat experiment 
due to the coupled propagation of magnetic and phononic excitations. 

Most important, this model predicts a specific two-bump 
counter-propagating temperature  profiles for certain material parameter range.
It would by particularly interesting to find materials with appropriate 
parameters in order to observe such a behavior in 
a dynamic heat propagation experiment, e.g. direct evidence of ballistic 
transport by spinons in spin-1/2 Heisenberg chain compounds.
Last but not least, althought the proposed model was motivated by the ballistic 
magnetic transport coupled to the diffusive phonon one in spin chains, 
it can be applied to any two-component system 
with coupled advection-diffusion transport. 

\bigskip
\section{Acknowledgments}
This work was supported by the
Deutsche Forschungsgemeinschaft through Grant HE3439/13.
X.Z. acknowledges fruitful discussions with Profs. P. van Loosdrecht and 
C. Hess on existing and future dynamic heat experiments.


\begin{thebibliography}{99}

\bibitem{hess}
C. Hess, Phys. Rep. {\bf 811}, 1 (2019).

\bibitem{ballistic}
N. Hlubek, P. Ribeiro, R. Saint-Martin, A. Revcolevschi, G. Roth, 
G. Behr, B. B\"uchner, C. Hess, Phys. Rev. B{\bf 81}, 020405 (2010).

\bibitem{znp}
X. Zotos, F. Naef and P. Prelovsek, 
Physical Review B {\bf 55}, 11029 (1997).


\bibitem{parker}
W. Parker, R. J. Jenkins, C. P. Butler, and G. L. Abbott, 
J. Appl. Phys. {\bf 32}, 1679 (1961).

\bibitem{flash} 
M. Montagnese, M. Otter, X. Zotos, D.A. Fishman, N. Hlubek, 
O. Mityashkin, C. Hess, R. Saint-Martin, S. Singh, A. Revcolevschi, 
P.H.M. van Loosdrecht, Phys. Rev. Lett. {\bf 110}, 147206 (2013).

\bibitem{ghd}
O.A. Castro-Alvaredo, B. Doyon and T. Yoshimura 
Phys. Rev. X{\bf 6}, 041065 (2016);
B. Bertini, M. Collura, J. De Nardis and M. Fagotti, 
Phys. Rev. Lett. {\bf 117}, 207201 (2016).

\bibitem{sw}
D.J. Sanders and D. Walton, Phys. Rev. B {\bf 15}, 1489 (1977). 

\bibitem{ott}
A. V. Sologubenko, K. Giann\`o, H. R. Ott,
A. Vietkine and A. Revcolevschi, Phys. Rev. B{\bf 64}, 054412 (2001).

\bibitem{cardano}
%G. Cardano, Artis magnae, sive de regulis algebraicis. Lib. unus. Qui  
%\& totius operis de arithmetica, quod opus perfectum inscripsit, est in ordine 
%decimus. (Ars Magna.) Johann Petreius, Nuremberg, 1545; 
Abramowitz, Milton; Stegun, Irene A., eds. Handbook 
of Mathematical Functions with Formulas, Graphs, and Mathematical Tables, 
Dover (1965), chap. 22 p. 773.

\end{thebibliography}
\end{document}